\documentclass[prd, showpacs,12pt]{revtex4-1} 

\usepackage{hyperref}
\usepackage{amsfonts, amsmath, amssymb}

\usepackage{geometry} 
\usepackage{slashed} 
\usepackage[pdftex]{graphicx}

\newcommand{\nn}{~,\nonumber\\}

\begin{document}

\title{Quantum electrodynamics of spin $3/2$}
\author{Konstantin G. Savvidy}
\affiliation{Physics Department, Nanjing University, Hankou Lu 22, Nanjing, P.R. of China}
\begin{abstract} 
Electromagnetic interactions of the spin 3/2 particle are investigated while allowing the propagation of the transverse spin 1/2 component present  in the reducible Rarita-Schwinger vector-spinor. This is done by allowing a more general form for the mass term, while leaving the kinetic terms untouched. We find that the interaction is consistent and does not lead to superluminal propagation for a range of the mass parameter where the spin 1/2 component is lighter than the spin 3/2 component, in contrast to the traditional value whereby the spin 1/2 component is removed by making it infinitely massive. The hyperbolicity condition is found to be independent of the magnitude of the electromagnetic field, and the canonically quantized field is constructed to be causal. We then provide appropriate projection operators which reproduce the spin-sum expressions and enable the construction of a physically acceptable propagator. Finally, we suggest a scheme for extension of the Standard Model with Rarita-Schwinger multiplets and identify the muon as the spin 1/2 component of the new multiplet which in addition contains a new, heavy partner of the muon of spin 3/2. The scattering cross-section $e^+,e^- \rightarrow \mu^+, \mu^-$ is calculated within this theory, compared to QED, and finally a bound on the new particle's mass is obtained from precision electroweak measurements.
\end{abstract} 


\maketitle
\tableofcontents

\setlength{\arraycolsep}{2pt}

Particle theories with particles of spin higher than one have held the attention of researchers with the promise for new physics and continue to do so. The Standard Model is built primarily out of spin 1/2 matter and spin 1 gauge fields. Einsteinian gravity, while containing a spin 2 graviton, cannot be considered a particle physics theory despite the efforts of generations of physicists. The Standard Model was constructed with the use of lowest spin fields, and only the simplest possible interactions are present. Due to historical coincidence, the Standard Model was also deemed to be a triumph of quantum field theory in its orthodox understanding over the much more exotic possibilities such as Regge theory, S-matrix theory and so on. Also due to historical circumstance, much of the early work on higher-spin field theory predates the acceptance and triumph of the Standard Model (SM), therefore it was deemed necessary to largely abandon those efforts. Only string theory, an outgrowth of the 60's heterodoxy continues to thrive, precisely because of its incorporation of gravity and hence higher-spin interactions. This has led to yet another perception, namely that any higher spin theory would be more similar to gravity, than the Yang-Mills gauge theory. In addition, two relevant no-go theorems have been proven, one by Coleman and Mandula \cite{Coleman:1967ad} stating that Lorentz symmetry cannot be extended, and a second one by Weinberg and Witten \cite{Weinberg:1980kq} which excludes the existence of conserved tensor currents, under certain conditions. 

Nevertheless, an interacting theory of higher-spin gauge fields, satisfactory at least at the classical level has been constructed by G. Savvidy \cite{Savvidy:2005fi,Savvidy:2005zm,Savvidy:2005ki}. The theory is free of ghosts \cite{Savvidy:2009zz}, and does not contain higher derivative non-renormalisable terms. The interactions are entirely of power-counting renormalizable cubic and quartic vertexes of the Yang-Mills type, and are dictated by a peculiar new extended gauge invariance principle. Moreover, the higher-rank tensor nature of the fields allows, following Deser, Jackiw and Templeton \cite{Deser:1982vy,Deser:1981wh} and Schonfeld \cite{Schonfeld:1980kb}, for mass terms of the Chern-Simons current type avoiding the need for the spin 0 Higgs particle \cite{Savvidy:2010bk} in four space-time dimensions.

Also recently, certain parts of open string amplitudes have been identified with those amplitudes in the tensionless string field theory \cite{Savvidy:2008ks}. Presence of higher spin matter and gauge particles is a generic feature of string theory and sharply contrasts with the generic prediction of Kaluza-Klein-type theories with partners of the same spin and supersymmetry with partners of spin lower by 1/2. Moreover because quantum numbers of any new particles are highly constrained, for example by the requirement of anomaly cancellation, it has been argued \cite{Cheng:2002ab,Datta:2005zs} that it may be very difficult to measure the spin of such partner particles by direct measurement. Therefore, it is important to develop quantitative theories which can distinguish between supersymmetric, Kaluza-Klein and higher-spin partners.

In light of these developments, it becomes urgent to develop phenomenologically realistic extensions of the Standard Model, incorporating the new gauge theory of higher spin. On this path, the major obstacle is the obscure status of interacting half-integer fermionic matter fields. On the one hand, satisfactory free theories of higher-spin fermions are known \cite{fierz,fierzpauli,Singh:1974qz,fronsdal}. On the other hand, the extended gauge invariance principle allows to fix most of the ambiguity in the interactions of matter with gauge fields. In the present work  the details of interactions of spin 3/2 matter field with spin 1 abelian gauge field are considered. This problem is also interesting in its own right, without reference to the new theories of higher-spin gauge fields.

The equations analysed in this paper are almost exactly those of the standard Rarita-Schwinger spin 3/2 theory \cite{Rarita:1941mf}. The difference is that we relax the Pauli-Fierz constraints, which are necessary to get rid of the reducible spin 1/2 components. The Rarita-Schwinger vector-spinor contains two spin 1/2 irreducible components, the undesirable longitudinal $ p_\mu \, \psi^\mu$ and the physical transverse $\gamma_\mu \, \psi^\mu$. At the level of the lagrangian and the equations of motions these two components can and do mix. We then find in Sec. \ref{sec:modes} that the theory contains one spin 3/2 and one spin 1/2 positive-norm physical particle, together with their antiparticles, while the other, approximately longitudinal spin 1/2 component does not propagate, i.e. does not go on-shell for any value of the momentum. The kinetic terms are fixed and are of the Rarita-Schwinger type. The mass term of this theory,  $m \, (\delta^\mu_{~\nu} - z \, \gamma^\mu \, \gamma_\nu)$ contains the only free parameter, which we call $z$, and which defines the ratio of the masses of the two physical particles as $M_{1/2}/M_{3/2} =  \frac {1-4z}{2-2z}$. 

The theory is then investigated as a function of this free parameter. The value of the parameter that corresponds to the standard choice $z=1$, whereby the spin 1/2 particle becomes infinitely massive is found to be untenable due to the state becoming longitudinal in that limit, and secondly because of the appearance of superluminal propagation, as found by Velo and Zwanziger \cite{Velo:1969bt,Velo:1970ur}. For most of the allowed range of the parameter, the spin 1/2 particle is lighter than the spin 3/2 particle, and this is also the phenomenologically preferred arrangement. We show in Sec. \ref{sec:super} that in the allowed range of the parameter superluminal propagation does not occur, leading to a quantum mechanically consistent theory with vanishing field anti-commutator at space-like distances (Sec. \ref{sec:field}). 

The propagator of the theory is found in Sec. \ref{sec:prop} by inverting the kinetic operator in momentum space. In order to write it in a usable form it is necessary to build up the projection operators to the separate physical subspaces; this is achieved by means of new projection operators, which is a linear combination of the well-known projection operators which separately project onto the transverse and longitudinal spin 1/2 subspaces. Our projectors therefore implement the spin-sums of the physical solutions found in Sec \ref{sec:modes}. Further, we make use of the appropriate parity operator in order to correctly propagate the positive (resp. negative) frequency solutions forward (resp. backward) in time. A comparison is made with those expressions found in the literature: in contrast to assorted spin 1/2 contributions in the standard theory we find only a purely longitudinal non-propagating contribution which may or may not contribute to any S-matrix elements. To clarify this issue it is necessary to compute matrix elements for some process with internal fermion lines, for example for Compton scattering and this will be investigated elsewhere.

In Sec.  \ref{sec:field} we consider the canonical quantization of the theory and present the second quantized field of the theory. The field anti-commutator is then computed and shown to vanish at space-like separations. The momentum space propagator is compared with 
the expression for the two-point function, and agreement is found. 

Calculations of Compton scattering with spin 3/2 particles in the intermediate state  arise in the phenomenological theory of photon-nucleon scattering \cite{Peccei:1969sb,Benmerrouche:1989uc,Pascalutsa:1995vx}, whenever the energy of the impinging gamma quantum $\approx 200-300 MeV$ is in the region of the $\Delta^+$ resonance. 
This remains an active subject of research \cite{Pascalutsa:1999zz, Pascalutsa:2006up, FernandezRamirez:2008tu, Lensky:2009uv}, and
our theory has the potential to bring clarity to some of the long-standing paradoxes and inconsistencies encountered in the investigation of this phenomenon, and this will be explored in a subsequent work. Specifically, we propose to describe the proton and the $\Delta^+$ as two different reducible components of a single Rarita-Schwinger multiplet, both particles being really different states of the same physical system composed of the same three quarks. Physical attractiveness of the unified description of the proton and $\Delta^+$ is  particularly convincing evidence in favor of allowing a spin 1/2 particle to propagate as part of the reducible Rarita-Schwinger vector-spinor representation. It is known \cite{Niederle:2004bw} that some other excited baryon states can also be consistently described in terms of multiparticle equations, specifically as pairs of opposite parity.

The formal developments in the first six Sections of the paper enable practical calculations of many relevant amplitudes and matrix elements. This is demonstrated in Sec. \ref{sec:anni} on the example of  basic $e^+,e^- \rightarrow \mu^+, \mu^-$ scattering. Here, the spin 1/2 physical particle of the Rarita-Schwinger multiplet is identified with the ordinary muon. It is verified that in the limit where the mass of the spin 3/2 component goes to infinity this matrix element for this process tends to the standard result in QED. 

The theory, understood as an extension of QED, predicts the existence of a spin 3/2 partner of the muon, with electromagnetic interactions completely specified by the present theory. When the mass of the new particle, smuon (for spinning), is taken to be around 1TeV,  the cross-section of the $e^+, e^- \rightarrow \mu^+, \mu^-$ scattering is expected, although we do not attempt a calculation of the cross-section by including the contributions from the Z-boson, to be around 1.5\% lower than that in the SM in the 160-207 GeV center-of-mass energy range. At the end of Sec. \ref{sec:anni} we consider whether precision electroweak measurements at LEP-II do not exclude deviations at such level, and in fact find that they favor this possibility slightly, the average ratio of the measured di-muon cross-section to the SM value is
less than one by $-1.98\% \pm 1.39\%$. 
In summary, although the magnitude of the deviation of the experimental measurement is slighly greater than our initial guess, caution is warranted due to marginal statistical significance. The best that can be said is that the data do not exclude the conjectured identification of the muon at this point. At higher energies, the modifications due to the present theory are more drastic and susceptible to measurement in current and future experiments.

\section{The Rarita-Schwinger field.}
\label{sec:rs}
There exists adequate formulation in the literature for the free equations of massive and massless fermions \cite{fierz,fierzpauli, Singh:1974qz, fronsdal}. Subsequently, it was discovered by Velo and Zwanziger \cite{Velo:1969bt,Velo:1970ur} that whenever the massive particle is coupled to the electromagnetic field, there exist solutions with superluminal propagation. Nevertheless, some calculations are unaffected by such difficulties in the formal theory, for example in the same
issue of Physics Review as the Rarita and Schwinger article there appeared an analysis by Kusaka \cite{Kusaka:1941} ruling out the possibility that the neutrino is a spin 3/2 particle described by the newly found equations. 

Before proceeding to give a proper solution to the problem of superluminal propagation in Sec. \ref{sec:super}, we present the preliminaries.
The general form of the spin $3/2$ lagrangian is
$$
L = \bar{\psi}_\mu ~ [D^\mu_{~\nu} - m\, \Theta^\mu_{~\nu}]~ \psi^\nu~,
$$
where the kinetic term $D$ is strictly linear in momentum and $\Theta$ is a mass term independent of momentum. Both are Lorentz tensors of the second rank:
\begin{eqnarray}
\label{eq:kin}
D^\mu_{~\nu} &=& (\gamma^\rho \, p_\rho)\, \delta^\mu_{~\nu} - 
(\gamma^\mu \, p_\nu +  \gamma_\nu \, p_\mu)
+ (\gamma^\mu \, \gamma^\rho \, p_\rho \, \gamma_\nu) \nn
 \Theta^\mu_{~\nu} &=& \delta^\mu_{~\nu} - z \, \gamma^\mu \, \gamma_\nu~.
\end{eqnarray}
All terms allowed by criteria mentioned are present in these tensors, however relative coefficients in $D^\mu_{~\nu}$ have been already tuned to the Rarita-Schwinger-Hagen-Fronsdal-Chang form.
The kinetic term is more or less unique up to contact transformations, a well flagged point in the existing literature. There is however some arbitrariness in the mass term parameter $z$.
The arbitrariness had been previously dealt in the literature by the following considerations. The vector-spinor field $\psi^\mu$ is a reducible representation of the Lorentz group. Having 16 components, it contains a $4+4=8$ degrees of freedom necessary to describe the spin 3/2 particle and its antiparticle. The remaining components fall into two Dirac spinors with 4 components each. 
Pauli and Fierz impose the conditions  $ p_\mu \, \psi^\mu = 0$ and $\gamma_\mu \, \psi^\mu=0$ by hand, while Rarita and Schwinger achieve this as a consequence of the equations of motion following from the lagrangian above with $z=1$.

\section{Propagating modes}
\label{sec:modes}
We proceed to investigate the solutions of the equations of motion $[D^\mu_{~\nu} - m\, \Theta^\mu_{~\nu}]~ \psi^\nu = 0$, while keeping $z$ as a free parameter.

The solutions with spin 3/2 do not depend on the parameter $z$, and are well-known. With gamma matrices in the Weyl representation, fourth component as time, vector polarization vectors 
$E_1=\tfrac {1}{ \sqrt{2}} \, (1,i,0,0)$,  
$E_2=\tfrac {1 }{\sqrt{2}} \, (1,-i,0,0)$,
$E_3=(0,0,1,0)$,
$E_4=(0,0,0,1)$, and a basis of Dirac spinors 
$$
U_1={\begin{pmatrix}1\\ 0\\ 1\\ 0 \end{pmatrix} }~, 
~~ U_2={ \begin{pmatrix}0\\ 1\\ 0\\ 1  \end{pmatrix} }~,~~
V_1={ \begin{pmatrix}1\\ 0\\ {-}1\\ 0  \end{pmatrix} }~,
 ~~ V_2={ \begin{pmatrix}0\\ 1\\ 0\\ {-}1  \end{pmatrix} }~.
$$
we may build up the positive frequency solutions with $\omega=m$ in the rest frame as
\begin{eqnarray}
\label{eq:e4}
u_4(0, \, +3/2)  &=& E_1 \otimes U_1\nn
u_4(0, \, +1/2)  &=& \tfrac{1}{\sqrt{3}} \,  E_1 \otimes U_2 - \tfrac{2}{\sqrt{3}} \,   E_3 \otimes U_1\nn
u_4(0, \, -1/2)  &=&  \tfrac{1}{\sqrt{3}} \,  E_2 \otimes U_1 + \tfrac{2}{\sqrt{3}} \,   E_3 \otimes U_2\nn
u_4(0, \, -3/2)  &=& E_2 \otimes U_2
\end{eqnarray}
The solutions are orthogonalized, with respect to each other, and in addition made to diagonalize the Rarita-Schwinger  spin operator
$\Sigma_3 =  \tau^{12} \otimes  \openone + \openone \otimes \sigma^{12} $, 
where $\tau$ and $\sigma$ are generators of rotations for ordinary vectors and spinors respectively. Thus, the solutions fill out the spin multiplet appropriate for a fermion of spin 3/2, and mass $M_{3/2}=m$.

Corresponding solutions with negative frequency $\omega=-m$ can be obtained by replacing $e$ for $p$ in \eqref{eq:e4}:
\begin{eqnarray}
\label{eq:p4}
v_4(0,+3/2)  &=& E_1 \otimes V_1\nn
v_4(0,+1/2)  &=& \tfrac{1}{\sqrt{3}} \,  E_1 \otimes V_2 -  \tfrac{2}{\sqrt{3}} \,  E_3 \otimes V_1\nn
v_4(0,-1/2)  &=&  \tfrac{1}{\sqrt{3}} \,  E_2 \otimes V_1 +  \tfrac{2}{\sqrt{3}} \,  E_3 \otimes V_2\nn
v_4(0,-3/2)  &=& E_2 \otimes V_2.
\end{eqnarray}
These are appropriate for the spin 3/2 anti-fermion.

The operator $D^\mu_{~\nu} - m \, \Theta^\mu_{~\nu} $ is acting on the 16-component vector-spinors and as such is generally expected to have 16 eigenvalues and eigenvectors. Instead, we are interested in homogeneous solutions to the equations, and thus need to investigate the \emph{nullspace} of the operator for different values of momenta. Because we are interested in massive solutions, it is convenient to consider particles in the rest frame, $p=(0,0,0,\omega)$. The characteristic polynomial is of the \emph{twelfth} degree in frequency $\omega$:
\newcommand{\Det}{\mathop{\mathrm{Det}}}
$$\Det |D^\mu_{~\nu} - m \, \Theta^\mu_{~\nu}| = [m + \omega]^4 \, [m - \omega]^4 \, 
 [M+\omega]^2 \, [M - \omega]^2~ m^4\,(2 - 2 \, z)^4    ~,
$$
where $M= m \, \frac {1-4 \, z}{2-2\, z}$.
Thus, $4+4=8$ eigenvalues can be nullified by setting $\omega=\pm m$ as above. Another $4$ eigenvalues can be made equal to zero at $\omega=\pm M$, two for each sign. The corresponding solutions are appropriate for describing a physical particle and antiparticle of spin $1/2$ and mass $M_{1/2} = M = m \, \frac {1-4 \, z}{2-2\, z}$:
\begin{eqnarray}
\label{eq:e2}
u_2(0,+1/2)  &=& {1/\sqrt{3}} \,  \left(E_1 \otimes V_2 + E_3 \otimes V_1 - \tfrac{3}{ A}\, E_4 \otimes U_1 \right)\nn
u_2(0,-1/2)   &=& {1/\sqrt{3}} \,  \left(E_2 \otimes V_1 - E_3 \otimes V_2 - \tfrac{3}{ A}\,  E_4 \otimes U_2 \right)\nn
v_2(0,+1/2)  &=& {1/\sqrt{3}} \,  \left(E_1 \otimes U_2 + E_3 \otimes U_1 - \tfrac{3}{ A}\,  E_4 \otimes V_1 \right)\nn
v_2(0,-1/2)   &=& {1/\sqrt{3}} \,  \left(E_2 \otimes U_1 - E_3 \otimes U_2 - \tfrac{3}{ A}\,  E_4 \otimes V_2 \right)\nn
A &=& 1-1/z~.
\end{eqnarray}
The naive norm $\psi^\dagger_\mu \psi^\mu$ of the states above is $ 1-3/A^2$ and becomes negative for $A^2 < 3$. Thus, we tentatively obtain the allowed range of the parameter $z$ as $z \in [\frac 1 {(1 - \sqrt 3)}, \frac 1 {(1+ \sqrt 3)} ]$. 

The standard choice $z=1$ leads to the disappearance of the above states from the theory simply because the mass of those states goes to infinity  $M= m \, \frac {1-4z}{2-2z} \rightarrow_{z\rightarrow1} \infty $. 
On the other hand, the value $z=1$ certainly lies outside the allowed region, since at the same time $A \rightarrow 0$ and the solutions become purely longitudinal in this limit. One might suspect already on this basis that trying to get rid of the spin 1/2 component by sending its mass to infinity would lead to trouble. We shall confirm this conclusion by investigating the causality of propagation of the waveform in the presence of the electromagnetic field in Sec. \ref{sec:super}, and the second quantized field in Sec. \ref{sec:field}.

The remaining four-dimensional vector subspace 
does not go on-shell for any value of frequency, because the characteristic polynomial is of the twelfth degree and allows only 12 propagating modes in \eqref{eq:e4}, \eqref{eq:p4} and \eqref{eq:e2}. This is of course the result of the fine tuning inherent in the Rarita-Schwinger-Fang-Fronsdal kinetic operator $D^\mu_{~\nu} $.

This completes the construction of the 12 propagating modes corresponding to a spin 3/2 ($4+4=8$ modes) and a spin 1/2 ($2+2=4$ modes) physical particle  in the rest frame, and thus by boosting in any frame.

\section{Electrodynamics}
\label{sec:ed}
The interaction with an abelian gauge field can be introduced in the usual way by covariant derivative $p_\mu \rightarrow \pi_\mu=p_\mu -  A_\mu$. This ensures gauge invariance and conservation  of the current:
$$
j^\mu = \frac{\delta L}{\delta p_\mu} = 
\bar{\psi}_\nu \, \gamma^\mu \, \psi^\nu + 
\bar{\psi}_\nu \, \gamma^\nu \gamma^\mu \gamma_\rho \, \psi^\rho -
\bar{\psi}_\nu \, \gamma^\nu \psi^\mu - \bar{\psi}^\mu \gamma_\nu \, \psi^\nu~.
$$
The same structure also appears in the cubic electromagnetic interaction vertex of the theory:
$$
\Gamma^\mu_{a b\, \nu\,\rho} = \frac{\delta L}{\delta A_\mu ~\delta \bar{\psi}^\nu_a  ~ \delta \psi^\rho_b} = 
\gamma^\mu_{a b} \, \delta^\nu_{~\rho} + \gamma^\nu_{a c} \, \gamma^\mu_{c d}\, \gamma_{\rho \, d b}
- \gamma^\nu_{a  b} \, \delta^\mu_\rho -  \gamma_{\rho \, a  b} \,  \delta^{\nu\mu}
$$
The structure of this vertex is very special, if it is contracted with the same vector twice the result is always zero: $V^\mu_{\nu\,\rho} \, a^\nu\,a^\rho = 0$. Further, it is important to ensure that the vertex leads to an amplitude which vanishes for longitudinal photons. What we need is a more-or-less standard Ward identity for the inverse of the fermion propagator:
$$
-i\, k_\mu \, \Gamma^{\mu\nu}_{~\rho}(p+k,p) = S^{\nu}_{~\rho}(p+k)^{ -1} - S^{\nu}_{~\rho}(p)^{ -1}
$$
and is satisfied, at least at the tree level, due to conservation of current together with the minimal coupling principle. At this level, the inverse of the propagator can be taken as the kinetic operator itself.

The spatial integral of the zero component of the current is electric charge:
$$
Q = \int j^0 = \int \bar{\psi}_\nu \, \gamma^0 \, \psi^\nu + 
\bar{\psi}_\nu \, \gamma^\nu \gamma^0 \gamma_\rho \, \psi^\rho -
\bar{\psi}_\nu \, \gamma^\nu \psi^0 - \bar{\psi}^0 \gamma_\nu \, \psi^\nu
$$
which Velo and Zwanziger rewrite as 
$$
Q = \int \vec{\psi}^\dagger \vec{\psi} - (\vec{\alpha} \cdot \vec{\psi})^\dagger (\vec{\alpha} \cdot \vec{\psi})~,
$$
and they also note that the quantity does not depend on the zero component of the wavefunctions. Further, they comment that the quantity is required to be positive definite whereas the expression above is manifestly not. If the spin 1/2 component is projected out by means of $\gamma_\nu\, \psi^\nu=0$, then the quantity can be made positive. In our version of the theory, the negative value of $-1$ exhibited by the spin 1/2 solution is simply interpreted as being due to the particle being of opposite charge for the positive frequency solution. The positive frequency solutions for the spin 3/2 and spin 1/2 particle have opposite parity, such that the net result is that particles of the same charge have the same, rather than opposite parity.  After second quantization charge should be expressed as a sum over species charge multiplied by the number operator - the more important requirement being that energy become positive after filling the Dirac sea. The electric charge of the Dirac particle is equal to the difference in the number of positrons and electrons $Q = \hat{N}_v -  \hat{N}_u$, while in the case at hand it turns out to be 
$$
Q = (\hat{N}^{3/2}_u -  \hat{N}^{3/2}_v) - (\hat{N}^{1/2}_u -  \hat{N}^{1/2}_v)~.
$$
In the next section we look at the wavefront propagation in the presence of non-zero electromagnetic field which is coupled to the particle as described above. 

\section{Superluminal wavefronts at $z=1$}
\label{sec:super}

It has been long known that as soon as an electromagnetic field is turned on, there appear physical wavefronts propagating faster than the speed of light \cite{Velo:1969bt,Velo:1970ur}. We shall investigate this phenomenon as a function of the parameter $z$.
Although it is possible to determine the dispersion relations, i.e. the dependence of $\omega$ on $\vec{p}^2$ by directly examining the kinetic operator $D$, we will follow tradition \cite{Niederle:2004bw} and examine the issue only in the relevant spin 1/2 sector by writing down the equations in the presence of the electromagnetic field, contracted with $\gamma^\mu$ and $\pi^\mu$ respectively:
\begin{eqnarray*}
\label{eq:vz1}
2 \, (\gamma \cdot \pi \gamma - \pi) \cdot \psi + m\, (1+4z) \, \gamma \cdot \psi &=& 0\nn
i \, F_{\mu\nu} \, \gamma^\nu \,  \psi^\mu + m \, \gamma \cdot \psi + z \, m \gamma \cdot \pi \,  \gamma \cdot \psi + i F_{\mu\nu} \sigma^{\mu\nu} \gamma \cdot \psi&=& 0~.
\end{eqnarray*}
The two equations can be combined by introducing the electromagnetic dual field $ \tilde{F}$ into 
$$
m\, (1-z) \, \pi \cdot \psi  - m^2 \, z/2 \, (1-4z) \gamma \cdot \psi - i \, g \, \gamma^5 \, \gamma \cdot \tilde{F} \cdot \psi = 0
$$

Looking for superluminal wavefronts in the eikonal approximation $\psi^\mu = \epsilon^\mu \, \exp( i \, \tau \, n_\mu \, x^\mu)$ at $\tau \rightarrow \infty$ leads to
\begin{equation}
m\, (1-z) \, \tau n_\mu \, \epsilon^\mu -  m^2 \, z/2 \, (1-4z) \gamma_\mu \, \epsilon^\mu - 
i \, g\,  \gamma^5 \, \gamma_\mu \cdot \tilde{F^\mu_\nu} \cdot \epsilon^\nu = 0
\end{equation}

The analysis of Velo and Zwanziger applies to the case $z=1$, whereby one can always find Lorentz frames with negative norm polarization
$\epsilon^\mu = (0,0,0,1)$ as a solution, meaning ultimately that there are superluminal solutions in every frame.
For $z\neq 1$ the first term dominates in the limit $\tau \rightarrow \infty$, forcing transversality $ n_\mu \epsilon^\mu =0$ thus there are no superluminal wavefronts in the general case. In fact, when $z\neq 1$ hyperbolicity no longer depends on the local value of the electromagnetic field, as it should not, according to \cite{Ranada:1980xp}.

The conclusion is that the appearance of superluminal waves is a direct and unavoidable consequence of the aesthetic preference to deal only with the spin 3/2 component, getting rid of both longitudinal and transverse spin 1/2 components by imposing both $ p_\mu \, \psi^\mu=0$ and $\gamma_\mu \, \psi^\mu=0$. What we find is that when the electromagnetic field is turned on, the tachyonic nature of the spin 1/2 component in the prohibited regime $z\rightarrow 1$ reasserts itself. Instead, when the latter constraints are relaxed away from the point $z=1$, the hyperbolicity is assured.

\section{The Propagator}
\label{sec:prop}

The operator $D^\mu_{~\nu} - m \, \Theta^\mu_{~\nu}$ is non-degenerate except for the points in momentum space where the two physical particles go on-shell, thus there should be no difficulty in constructing the propagator.
In order to guess the inverse matrix $S^\mu_{~\nu}(p) = (D^\mu_{~\nu} - m \, \Theta^\mu_{~\nu})^{-1}$ it is then necessary to first obtain the correct form of the spin-sum expressions that go into the numerator. To do that, we need to express the projection operators (spin sums) as Lorentz covariant tensor expressions. The ingredients are the following set of projection operators:
\begin{eqnarray}
\label{eq:proj}
\Pi^{3/2} &=& \delta^\mu_{~\nu} - \tfrac 1 3 \, \gamma^\mu \, \gamma_\nu - 
                     \tfrac 1 {3p^2} \, (\slashed{p} \, \gamma^\mu \, p_\nu + p^\mu \, \gamma_\nu \, \slashed{p}) \nn
\Pi^{1/2}_{11} &=&  \tfrac 1 3 \, \gamma^\mu \, \gamma_\nu - \tfrac 1 {p^2} \,  p^\mu \, p_\nu +
                     \tfrac 1 {3p^2} \, ( \slashed{p} \, \gamma^\mu \, p_\nu + p^\mu \, \gamma_\nu \, \slashed{p}) \nn
 \Pi^{1/2}_{22} &=& \tfrac 1 {p^2} \, p^\mu \, p_\nu \nn
 \Pi^{1/2}_{21} &=&  \tfrac 1 { p^2} \,  ( p^\mu \, p_\nu  - \slashed{p} \, \gamma^\mu \, p_\nu) \nn
 \Pi^{1/2}_{12} &=&  \tfrac 1 { p^2} \,  ( \slashed{p} \, p^\mu \, \gamma_\nu  - p^\mu \, p_\nu )~.
\end{eqnarray}
This set of projection operators has the following meaning \cite{neuw, Benmerrouche:1989uc, Pascalutsa:1995vx}. $\Pi^{3/2}$ projects to the pure transverse spin 3/2 subspace, $\Pi_{22}$ to the longitudinal spin 1/2 subspace, and finally $\Pi_{11}$ to the transverse  spin 1/2 subspace. Lorentz invariance alone does not dictate that the two spin 1/2 subspaces will not mix in the lagrangian or the equations of motion. This necessitates the remaining two operators, $\Pi_{12}$ and $\Pi_{21}$ which respectively take vectors from longitudinal to the transverse space and the reverse.

Because the $\gamma_\mu \, \psi^\mu$ and the $p_\mu \, \psi^\mu$ mix, the propagating spin 1/2 mode is picked out by means of constructing an appropriate linear combination. We have constructed  a new set of mutually orthogonal and complete projection operators:
\begin{eqnarray}
\label{eq:myproj}
\Pi_3 &=& {\openone} - \Pi_B - \Pi_C = \Pi^{3/2}\nn
\Pi_B &=& \left( \Pi_{11} - \tfrac 1 A \, (\Pi_{21} + \Pi_{12}) + \tfrac 3 {A^2} \, \Pi_{22}\right) ~ \frac {A^2} {A^2 +3} \nn
\Pi_C &=& 
\left(  3 \, \Pi_{11} + A   \, (\Pi_{21} + \Pi_{12}) + A^2 \, \Pi_{22}\right) ~ \frac {1} {A^2 +3}
\end{eqnarray}
The operators $\Pi_3$, $\Pi_B$ and $\Pi_C$ have respectively 8, 4, and 4 unit eigenvalues. Note that exchanging $A$ for $-3/A$ exchanges $\Pi_B$ and $\Pi_C$. Also, in the $A \rightarrow \infty$ limit $\Pi_B$ coincides with the purely transverse $\Pi_{11}$, while $\Pi_C$ approaches the purely longitudinal $\Pi_{22}$.

Further, in order to restrict to the positive and negative energy subspaces separately, such as to propagate the corresponding states forward (resp. backward) in time, we define two additional projection operators for projecting to the parity odd and even eigenspaces:
\begin{eqnarray}
\label{eq:parity}
 \Pi^\pm(\Omega) &=& \frac 1 2 ~ \left({\openone} \pm \mathbb{P}  \right)\nn
\mathbb{P} &=& \left(\delta^\mu_{~\nu} - 2 \, \frac { p^\mu \, p_\nu} {p^2}\right)  \otimes  \frac {\slashed{p}} {\Omega}\nn
\mathrm{where}~ \Omega &=& \sqrt{ p^2},\,m, \, M.
\end{eqnarray}
It will be necessary to specify the value of $\Omega$ depending on the intended use, i.e. in the case of the propagators and spin sums it is more convenient to work with the on-shell value of $\Omega$, equal to mass.

We have checked explicitly that the space spanned by positive and negative energy solutions coincides
with the subspaces projected to by our projectors:
\begin{eqnarray}
\label{eq:spinsum}
 \sum_{s=-3/2}^{3/2} u_4(\mathbf{p}, \,s) ~\, \bar{u}_4(\mathbf{p}, \,s) &=&   {-} \Pi^+\, \Pi_3  \nn
 \sum_{s=-3/2}^{3/2} v_4(\mathbf{p}, \,s) ~\, \bar{v}_4(\mathbf{p}, \,s) &=&   \Pi^- \, \Pi_3  \nn
 \frac {2\,A^2} {A^2 +3} ~\sum_{s=-1/2}^{1/2} u_2(\mathbf{p}, \,s) ~\, \bar{u}_2(\mathbf{p}, \,s)   &=&  \Pi^- \, \Pi_B  \nn
 \frac {2\,A^2} {A^2 +3} ~\sum_{s=-1/2}^{1/2} v_2(\mathbf{p}, \,s) ~\, \bar{v}_2(\mathbf{p}, \,s)   &=&   {-} \Pi^+\,\Pi_B ~.
\end{eqnarray}
On the contrary, the subspace spanned by $\Pi_C$, is not collinear with the \emph{nullspace} of the kinetic operator  $D^\mu_{~\nu} - m \, \Theta^\mu_{~\nu}$ for any value of momenta. Thus, longitudinal, or any other  negative norm states close to it in the sense of being spanned by $\Pi_C$ do not propagate. We shall see that again by direct inspection of the propagator, below.

With these ingredients, our result for the propagator $i\,(D^\mu_{~\nu} - m \, \Theta^\mu_{~\nu})^{-1}$ is:
\begin{eqnarray}
\label{eq:prop}
-i\,S^\mu_{~\nu}(p) &=& \frac{2 \, m \, \Pi^+{(\Omega=m)}  }{p^2-m^2+i\,\epsilon} \, \Pi_3 \nonumber\\
&+&\frac{2\,M \, \Pi^-{(\Omega=M)}  }{p^2-M^2+i\,\epsilon} \, \frac {A^2 +3} {2\,A^2}\, \Pi_B + \,{\frac {2\,(M-2\,m)}{{3m}^{2}}} \, \Pi_{22} \nonumber\\
\end{eqnarray}
It is necessary to further elucidate the meaning of this 
expression and compare to those appearing previously in the literature \cite{neuw,Benmerrouche:1989uc,Pascalutsa:1995vx}.

First, in order to get the correct decomposition between positive and negative energy propagating modes we must use parity projection operators which correctly take into account the vector-spinor nature of the Rarita-Schwinger wavefunction. The operators as appearing above in \eqref{eq:prop} have mass substituted in the denominator, rather than $\sqrt{p^2}$, just like in Dirac theory. Existing literature \cite{neuw,Benmerrouche:1989uc,Pascalutsa:1995vx} did not make use of the correct vector-spinor parity operator in the propagator, simply because it is unnecessary in the spin 3/2 term and can be safely replaced by its Dirac spinor equivalent $(1+\slashed{p}/m)$, while we find it crucial if the physical spin 1/2 component is to propagate properly.

Second, the purely longitudinal mode corresponding to $\Pi_{22}$ contributes without ever going on-shell, with an amplitude independent of momentum in \eqref{eq:prop}. This does not lead to disastrous consequences, even though the interaction vertex  allows the coupling to such purely longitudinal states. Both here and in standard QED it is necessary to show that the S-matrix vanishes with longitudinal polarization states; the difference is that here they never propagate on-shell and at most contribute to the off-shell Greens functions but not the S-matrix. This is seen most easily in the LSZ formalism: the S-matrix elements are read off from the poles in the unamputated Greens function, but propagator above does not contain poles corresponding to the longitudinal states.

Finally, the limit of the second part as $M \rightarrow \infty$ is nonzero. Thus, even as the spin 1/2 component becomes infinitely massive, its influence does not go away even in that limit. The presence of those terms led Pascalutsa and Scholten\cite{Pascalutsa:1995vx} to comment that the extra pieces must be due to some high-lying resonances, with the momentum dependence of the propagator being washed out due to the high mass of the physical particle. 

In our theory, the spin 1/2 component is interpreted as the progenitor, and most of the allowed region for its mass lies below that of the spin 3/2 component; certainly the limit $M \rightarrow \infty$ is illegal in the present theory. 
On the contrary, the case $M << m$ is the important physical limit to consider in extensions of the  Standard Model, if one wishes to identify some known particle as the spin 1/2 component of the multiplet and argue that the spin 3/2 component is much heavier.

This completes the construction of a propagator with poles corresponding to physically propagating particles and only them, and usable for calculations of scattering amplitudes. 

\section{Causal Quantum Field}
\label{sec:field}

The propagating Rarita-Schwinger field is built up as follows:
\begin{eqnarray}
\label{eq:qft}
\psi^\mu (\mathbf{x}) &=&
\frac 1 {8\pi^3}\, \int \!\!\int \!\!\int d^3 \mathbf{p} \, \exp{( i \, \mathbf{p}\, \mathbf{x})} \nonumber\\
&&\left(
\sum_{s=-3/2}^{3/2} \sqrt{ \frac{m}{ { E_{\mathbf{p}}(m)}  } } 
\left[ a( \mathbf{p} , 3/2, s) \, u_4(\mathbf{p}, \,s) + {b}^{\dagger}( -\mathbf{p}, 3/2, s) \, v_4 ( -\mathbf{p} ,-s) \right] \right. \nonumber\\
&+& \left. \sum_{s=-1/2}^{1/2}\sqrt{ \frac{M}{ { E_{\mathbf{p}}(M)}  } } 
\left[ a( \mathbf{p}, 1/2, s) \, u_2 ( \mathbf{p}, s)+{b}^{\dagger}( -\mathbf{p}, 1/2, s) \, v_2 ( -\mathbf{p} , -s) \right]
 \right) ~,\nonumber\\
 E_{\mathbf{p}}(\Omega) &=& \sqrt{ \mathbf{p}^2 + \Omega^2}\nn
\end{eqnarray}
with the standard anti-commutation relations for all operators:
$$
\{ \,a(\mathbf{p}, l, r) \,, \,a^{\dagger}(\mathbf{q} , l, s) \,\} = \{ \,b(\mathbf{p}, l, r) \,, \, b^{\dagger}(\mathbf{q},l,s)  \,\} = 
(2\pi)^3 \;\delta^{(3)} (\mathbf{p} - \mathbf{q}) \, \delta^{rs} ~~, ~~~~l=1/2, 3/2~.
$$
Finally, the equal-time field anti-commutators then follow
\begin{eqnarray}
\label{eq:comm}
\{ \, \psi^\mu_a ( \mathbf{x}) \, , \, \bar\psi^{\nu}_b ( \mathbf{y}) \, \}  = ~~~~~~~~~~~~~~~~~~~~~~~~~~~~~~~~~~~~~~~~~~~\nonumber\\
  {} - 2\,m\,\left(\Pi^+\, \Pi_3 ~  \Delta_+^m \,( \mathbf{x} - \mathbf{y})
- \Pi^- \, \Pi_3   ~  \Delta_+^m \,(  \mathbf{y} - \mathbf{x})\right) \\
 {} +  2\,M\, \frac {A^2 +3} {2\,A^2} \left(\Pi^-\, \Pi_B ~  \Delta_+^M ( \mathbf{x} - \mathbf{y})
 - \Pi^+ \, \Pi_B  ~   \Delta_+^M (  \mathbf{y} - \mathbf{x}) \right) \nn
\{ \, \psi^\mu_a ( \mathbf{x}) \, , \, \psi^{\nu}_b ( \mathbf{y}) \, \} = 
\{ \, \bar\psi^{\mu}_a ( \mathbf{x}) \, , \, \bar\psi^{\nu}_b ( \mathbf{y}) \, \} = 0 ~,
\end{eqnarray}
where we introduced the usual massive scalar propagator as
$$
\Delta^\Omega_+(\mathbf{x},t) = \frac 1 {8\pi^3} ~ \int \frac {d^3 \mathbf{p} } { {2\, E_{\mathbf{p}}(\Omega)}}
~e^{ i \, \mathbf{p}\, \mathbf{x} - i\,E_{\mathbf{p}}(\Omega)\,t}
$$

The scalar propagator is even in coordinates at space-like separation, $\Pi_B$ and $\Pi_3$ are also both even. The parity projectors $\Pi^\pm$ have mixed properties under reflection.
The net result is that the field anti-commutators vanish for space-like separation.

It is not hard to check, by using the Feynman prescription for the pole in $p_0$ \eqref{eq:prop}, that the first two terms of the momentum space propagator \eqref{eq:prop} reproduce, after integration over  $p_0$ the two-point function above \eqref{eq:comm}. The third term in  \eqref{eq:prop} does not contain a pole and therefore does not contribute other than a contact term to the two-point function. Whether it can be ignored or whether it makes contributions to processes with internal fermion lines is an open question.

Finally, for completeness, we provide the expressions for charge and energy:
\begin{eqnarray}
Q &=& \int d^3 \mathbf{p} ~ \left[(\hat{N}^{3/2}_u -  \hat{N}^{3/2}_v) - (\hat{N}^{1/2}_u -  \hat{N}^{1/2}_v) \right]\nn
E &=& \int d^3 \mathbf{p} ~  \left[E_{\mathbf{p}}(m)(\hat{N}^{3/2}_u + \hat{N}^{3/2}_v) + E_{\mathbf{p}}(M) \,  (\hat{N}^{1/2}_u + \hat{N}^{1/2}_v)\right]~,
\end{eqnarray}
where 
\begin{eqnarray*}
\hat{N}^{3/2}_u = \sum_{s=-3/2}^{3/2} a^{s\dagger}_\mathbf{p} \, a^s_\mathbf{p} & \quad,\quad
\hat{N}^{3/2}_v = \sum_{s=-3/2}^{3/2} b^{s\dagger}_\mathbf{p} \, b^s_\mathbf{p} ~,\\
\hat{N}^{1/2}_u = \sum_{s=-1/2}^{1/2} a^{s\dagger}_\mathbf{p} \, a^s_\mathbf{p} & \quad,\quad
\hat{N}^{1/2}_v = \sum_{s=-1/2}^{1/2} b^{s\dagger}_\mathbf{p} \, b^s_\mathbf{p} ~.\\
\end{eqnarray*}

\section{Beyond the Standard Model}
\label{sec:anni}

Motivated by the existence of the extension of gauge theory to higher spins, we would like to extend the Standard Model matter with extra vector-spinor multiplets, enriching the SM with new particles.
Two schemes are possible in this regard. 

The first, basic possibility  \cite{Savvidy:2005zm} is that every SM fermion gets a spin 3/2 partner multiplet, which appears to require in addition to the spin 3/2 particle to include another, heavier spin 1/2 particle. This amounts to 4 additional multiplets per SM generation, a rather large number of particles.

The second, more economic possibility is that the enlargement occurs by generation, i.e. only in the second and third generations. The muon is then identified as the spin 1/2 component of the spin 3/2 multiplet, thus adding only two extra leptonic particles in the second generation: the heavy spin 3/2 muon and a heavy spin 3/2 muonic neutrino. The second of these, the spin 3/2 heavy muonic neutrino, being the lightest, is identified as the dark matter candidate particle. One might even speculate that the third generation leptons are part of a spin 5/2 multiplet.

In the following we compute the pair annihilation amplitude $e^+, e^- \rightarrow \gamma \rightarrow \mu^+, \mu^-$ under the assumption that the muon is part of a spin 3/2 Rarita-Schwinger multiplet containing the new spin 3/2 lepton, which we call smuon. This is in analogy to the MSSM naming scheme, except "s" stands for spinning rather than super.

Pair annihilation amplitude involves expressions of rather large size, making computation  by the method of Feynman trace a little bit too heavy even with the help of software. By contrast, computing helicity basis amplitudes by constructing the wavefunctions explicitly, boosting them, and contracting with explicit gamma-matrices substituted into for the vertex does not present any problem. Worksheet in the {\textsc{Maple}} computing language is available upon request from the author. In this way, we have been able to obtain the pair annihilation matrix elements, for the $e^+, e^- \rightarrow \mu^+, \mu^-$ process. The calculations are verbose but straightforward; simplification results in the ultra-relativistic limit for the electron $m_e \rightarrow 0$. We present only the results, namely for
the differential cross-section
\begin{eqnarray*}
&&\frac {d\Sigma}{d\Omega} =
\frac { {\alpha}^{2}} { 3 \, E^2}\sqrt {1-{\mbox {{ (M/E)}}}^{2}} ~\nonumber\\
&&\frac 1 {48 \,{E^{2}{M}^{4}}}
\left[
 {\, \left( (3{M}^{2}-4\,E^{2}{r}^{2})^2- (4\,M\, r \left( r-1\right) E)^2 \right)  \left( E^2 - M^2
 \right) 
 \left( \cos \left( \theta \right)  \right) ^{2}}+\right. \nonumber\\
&& \left. 16\,E^{6}{r}^{4}-32\,{M}^{2} \left( r-1/2 \right) ^{2}{r}^{2}E^{
4}+16\,{M}^{4} 
\left( {r}^{3}-1/2\,{r}^{2}+7/4\,r-3/8 \right)  \left( r-3/2 \right) E^{2}+9\,{M}^{6} \right]
\end{eqnarray*}

and total cross-section
\begin{eqnarray*}
\label{eq:cs}
&&\Sigma_{\textrm{tot}} =
\frac {\pi \, {\alpha}^{2}} { 3 \, E^2}\sqrt {1-{\mbox {{ (M/E)}}}^{2}} ~\nonumber\\
&&\left[
{\frac {16}{9}}\,{\frac { \left( {{\it E}}^{4}+{M}^{4}-2\,{{\it E}}^{2}{M}^{2} \right) {r}^{4}}{{M}^{4}}}+{\frac {32}{9}}\,{
\frac { \left( -{M}^{2}+{{\it E}}^{2} \right) {r}^{3}}{{M}^{2}}}- 
{\frac {8}{9}}\,{\frac { \left( 2\,{{\it E}}^{2}-5\,{M}^{2}
 \right) {r}^{2}}{{M}^{2}}}-4\,r+\frac 1 2\,{\frac {{M}^{2}+2\,{{\it E}}^{2}}{{{\it E}}^{2}}}
 \right]
\end{eqnarray*}
where $r$ is the presently unknown ratio of muon to spin 3/2 smuon mass $r = M/M_{3/2} = \tfrac {1-4z}{2-2z}$, and $M$ is the actual muon mass of $\sim 0.105 GeV$. Crucially, in the limit $r \rightarrow 0$ where the new spin 3/2 particle becomes infinitely heavy this reduces to the standard QED answer. Nevertheless, the cross-sections as they stand are not physically acceptable in the very high energy region because they grow as $E^2$. However, the higher-spin gauge theory of G. Savvidy contains other contributions to the process which are sub-leading with respect to the electromagnetic exchange considered here, due to the high mass of the new gauge bosons. We conjecture that the  s-channel contribution which regularizes the scattering cross-section at high energy comes from one of the spin 1 heavy gauge bosons contained in the higher-rank tensor gauge fields, for example the rank-3 vector field $A_{\mu\lambda\lambda}$.

\begin{figure}[ltbp]
\begin{center}
\includegraphics[scale=.5]{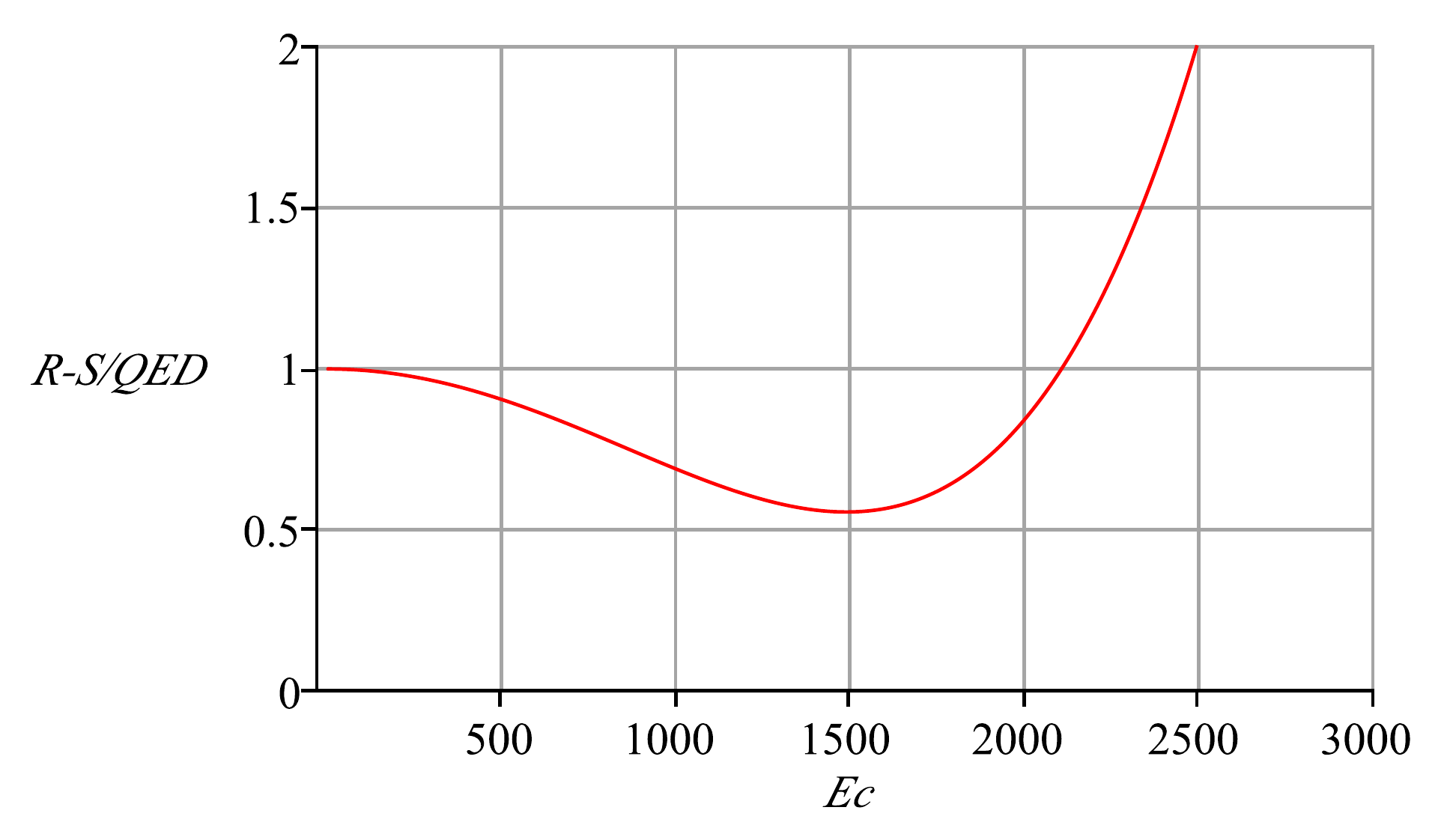}
\caption{{\small The ratio of the value of the total pair annihilation cross-section in the present theory to the standard QED value, for $r=10^{-4}$, as a function of c.m. energy in GeV.}\label{fig:ratio}}
\end{center}

\end{figure}

The ratio of the cross-section \eqref{eq:cs} to the standard QED expression is plotted in Fig. \ref{fig:ratio}.
The cross-section does not exceed the QED value immediately, instead the ratio drops below one quadratically with energy until reaching a minimum at approximately the energy scale associated with the heavy spin 3/2 partner.
Assuming the mass of the new particle around 1 TeV, the muon/smuon mass ratio is $r \approx 100MeV/1TeV \approx 10^{-4}$. The  new cross-section is about $1.5\%$ lower than QED prediction at LEP-II energies of 160-207 GeV.
For higher energy, the cross-section drops faster than QED, reaching a minimum at 1.4 TeV, then starts to grow drastically beyond 2 TeV.

\begin{figure}[htbp]
\begin{center}
\includegraphics[scale=.35]{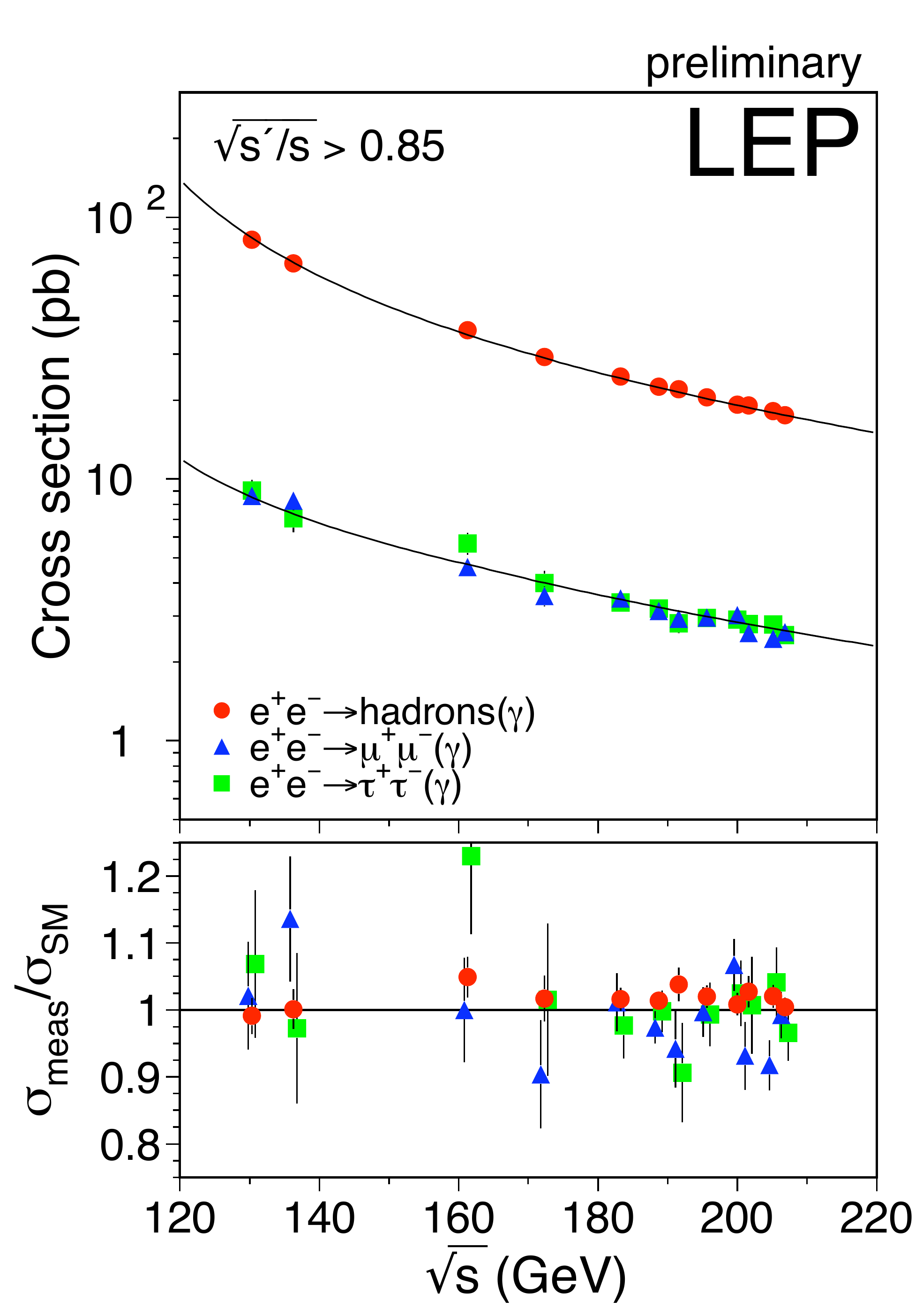}
\caption{Preliminary combined LEP results on the cross-sections for qq, $\mu^+\mu^-$ and $\tau^+\tau^-$ final 
states, as a function of centre-of-mass energy. The lower plot shows the ratio of the data divided by the SM value. 
(Courtesy of the LEP Electroweak Working Group.) \label{fig:ewwg}}
\end{center}
\end{figure}

The calculation does not include the contribution of the heavy neutral currents, i.e. the diagrams with the exchange of the $Z_0$. Nevertheless it is interesting to ask whether the precision electroweak data from LEP-II exclude such low value of the smuon mass.  The raw data was taken from the LEP-II precision measurements of the Electroweak Working Group \cite{ewwg}, which provides the di-fermion cross-sections combined from  ALEPH, DELPHI, L3, and OPAL.  Evident from the figure is that nearly all di-quark measurements are slightly above the SM prediction, while most di-muon cross-sections are lower. This general pattern persists also when one drills down from the combined to the individual detector data. We use the data from the high-luminocity runs from 160-207 GeV, weighted by the size of error bars. Alone, the di-muon weighted average cross-section ratio is slightly smaller than one: $-1.98\% \pm 1.39\%$, while the di-quark ratio is slightly above one: $+1.68\% \pm 0.33\%$. The error estimate ignores systematics, which are correlated between data points; the latter value for di-quark ratio most probably means that the pull from systematics had been upward. 

\begin{figure}[htbp]
\begin{center}
\includegraphics[scale=.45]{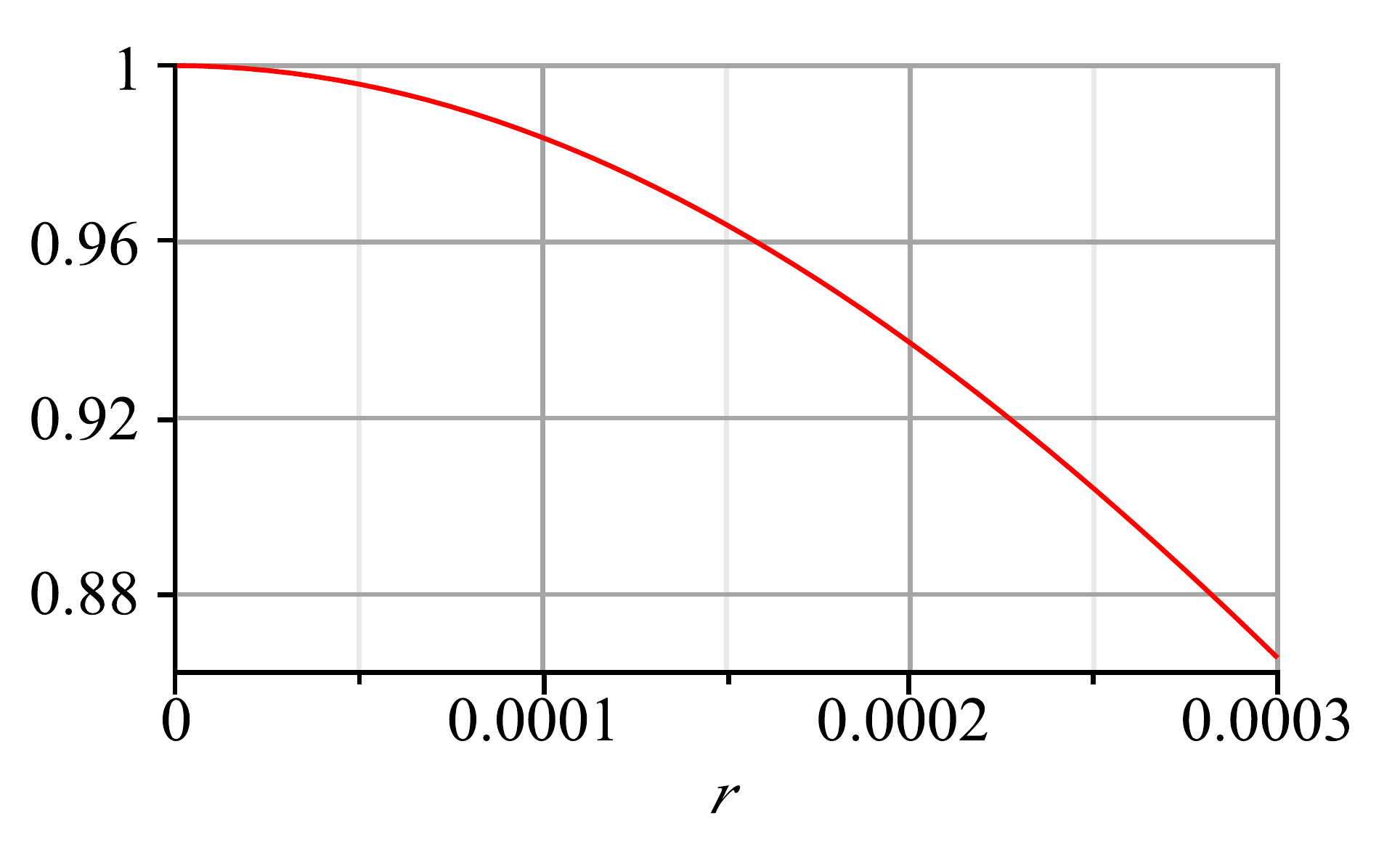}
\caption{The dependence of the new cross-section ratio to QED at Ec=200 GeV on the muon/smuon mass ratio $r$. \label{fig:massr}}
\end{center}
\end{figure}

The average value of the di-muon cross-section ratio is different from zero in the direction favorable for the new theory, but of a marginal statistical significance. 
The best that can be said therefore is that the data do not exclude the existence of the new smuon particle based on its influence on the di-muon cross-section, with lower bound on the mass of about  600 GeV, and preferred range of 800-1000 GeV (see Fig. \ref{fig:massr}).

\begin{acknowledgments} 
The author would like to thank Ian McArthur and Sergei Kuzenko of the University of Western Australia, Viatcheslav Mukhanov of the University of Muenchen, and Johan Bijnens and Torbj\"{o}rn Sj\"{o}strand of Lund University for hospitality and discussions, and Edna Cheung and George Savvidy for encouragement and many discussions.

\end{acknowledgments}

\vfill

\end{document}